\RequirePackage{fix-cm}
\documentclass[twocolumn,epjc3]{svjour3}  
\smartqed  
\RequirePackage{graphicx}
\usepackage[square,numbers,comma,sort&compress]{natbib}
\usepackage{tabularx} 
\usepackage{hyperref}
\usepackage{graphicx}%
\usepackage{multirow}%
\usepackage{lipsum}
\usepackage{amsmath,amssymb,amsfonts}%

\hypersetup{
    colorlinks=true,
    linkcolor=blue,
    filecolor=magenta,      
    urlcolor=cyan,
    citecolor=blue,
    }
\begin{document}

\title{Estimating centrality in heavy-ion collisions using Transfer Learning technique}





\author{Dipankar Basak\thanksref{e1,addr1, addr2}
        \and
        Kalyan Dey\thanksref{e2,addr1} 
}

\thankstext{e1}{e-mail: dipankar0001@gmail.com}

\thankstext{e2}{e-mail: kalyan.dey@buniv.edu.in (corresponding author)}


\institute{Department of Physics, Bodoland University, Kokrajhar, 783370, Assam, India \label{addr1}
           \and
           Department of Physics, Kokrajhar Govt. College, Kokrajhar, 783370, Assam, India \label{addr2}
}

\date{Received: date / Accepted: date}

\maketitle

\begin{abstract}
In this study, we explore the applicability of Transfer Learning techniques for estimating collision centrality in terms of the number of participants ($N_{\rm part}$) in high-energy heavy-ion collisions. In the present work, we leverage popular pre-trained CNN models such as VGG16, ResNet50, and DenseNet121 to determine $N_{\rm part}$ in Au+Au collisions at $\sqrt{s}=200$ GeV on an event-by-event basis. Remarkably, all three models achieved good performance despite the pre-trained models being trained for databases of other domains. Particularly noteworthy is the superior performance of the VGG16 model, showcasing the potential of transfer learning techniques for extracting diverse observables from heavy-ion collision data. 


\end{abstract}

\section{Introduction}\label{Introduction}
One of the main objectives of various heavy-ion collision (HIC) programs at different accelerator facilities is to scan the entire phase diagram of strongly interacting matter \cite{Busza2018} and to study the properties of the created matter. At ultra-relativistic beam energies, when two Lorentz-contracted nuclei collide, they produce an enormous energy density at the collision zone. The energy deposited during a collision is crucial in determining the post-collision state of matter and the relevant degrees of freedom. The amount of energy deposited depends strongly on the extent of overlap between the colliding nuclei. Lattice QCD calculations suggested that exceeding an energy density of 1 $\mathrm{GeV/fm^3}$ in collisions triggers a transition from a hadronic phase to a deconfined state of matter known as Quark-Gluon Plasma (QGP), where quarks and gluons are no longer confined within individual hadrons. Consequently, the nature, properties, and space-time evolution of the created matter in relativistic HIC depend strongly on collision geometry. Determination of centrality is often considered a top priority for analyzing experimental data on HIC. The concept of centrality is introduced to quantify the nuclear overlap and, hence, the initial geometry of heavy-ion collisions. Theoretically, collision geometry can be described by various proxies such as impact parameter ($b$), number of binary nucleon-nucleon collisions ($N_{\rm coll}$), and number of participating nucleons ($N_{\rm part}$). The mentioned quantities are not, however, directly measurable in experiments and are generally derived from the experimental data recorded during the collisions by using other quantities that scale approximately with them. For this purpose, the Glauber model \cite{Miller:2007} is often used. The Glauber model applies quantum-mechanical scattering theory to understand the interactions of composite systems, allowing the calculation of geometric quantities that can be connected to experimentally measurable observables such as charge particle multiplicity or forward energy carried by spectator nucleons, etc \cite{ALICE-PUBLIC-2018-003}.
 
The recent years have witnessed a surge in the application of machine learning (ML) and deep learning (DL) techniques across diverse scientific disciplines. In the field of high-energy nuclear physics, these methods have demonstrated significant success compared to conventional approaches \cite{Bass:1996,Pang:2016,Monk:2018,Li:2020,Du:2020,Tsang:2021,Mallick:2021,Song:2021,Kuttan:2021,Li:2021,Zhao:2022,arxiv.2111.15655,Shokr:2022,Xiang:2022,Zhang:2022,Saha:2022,Mallick:2022,Mallick:2023}. These data-driven techniques have used for various tasks in high-energy nuclear physics, including estimating the impact parameter of collisions \cite{Li:2020,Mallick:2021,Tsang:2021,Li:2021,Saha:2022,Zhang:2022,Xiang:2022}, analyzing flow observables \cite{Hirvonen:2023}, and even helping to identify the nature of the elusive QCD phase transition \cite{Du:2020}. Machine learning algorithms like Convolutional Neural Networks (CNN) and Light Gradient Boosting Machine (LightGBM) were used successfully by Li et al. \cite{Li:2020,Li:2021} to estimate impact parameter for Au+Au collisions at intermediate energies (0.2-1.0 GeV/nucleon) and Sn+Sn collisions at the beam energy of 270 MeV/nucleon respectively. With reasonable accuracy, both studies could predict the impact parameter. While achieving good accuracy, their results were sensitive to the parameter set of the event generator used for simulating heavy-ion collisions. Xiang et al. \cite{Xiang:2022} used the energy spectra for final-state charged hadrons to calculate the impact parameter for Au+Au collisions at $\sqrt{s} = 200$ GeV using multi-layer perceptron (MLP) and convolutional neural network (CNN) models. Both networks performed well in estimating impact parameter for semi-central and semi-peripheral collisions, with the CNN slightly outperforming the MLP model. In ref. \cite{Zhang:2022}, constrained molecular dynamics simulations for $^{124}\rm Sn+^{124}\rm Sn$ collisions at low-intermediate (50 - 100 MeV/nucleon) incident energies were utilized to predict impact parameter using the convolutional neural network (CNN). The study demonstrated that this approach outperforms traditional methods in terms of accuracy across all impact parameter ranges, particularly for central collisions. Elliptic flow ($v_2$) in heavy-ion collisions at the RHIC and LHC energies was estimated by Mallick et al. \cite{Mallick:2022,Mallick:2023} for Pb-Pb collisions at $\sqrt{s} = 5.02$ TeV using DNN model. The results showed a good agreement between the model prediction and simulated true value in both cases. In our previous work \cite{Basak2023}, we demonstrated the precise determination of collision centrality in terms of $N_{\rm part}$ through the effective utilization of Deep Neural Networks (DNN and CNN). In that investigation, the studied deep learning models exhibited remarkable proficiency in predicting $N_{\rm part}$, with CNN slightly outperforming DNN. The accuracy of the Deep Learning models was found to be particularly high for semi-central and central collisions, compared to peripheral collisions. Moreover, these DL models exhibited robustness, implying their ability to maintain satisfactory performance even with slight alterations in collision energy or model parameters at the event generator level.\\

In the present investigation, an attempt has been made to explore the possibility of determination of $N_{\rm part}$ employing the Transfer Learning technique within the framework of Deep Learning. Transfer Learning, a powerful technique within Deep Learning, has gained significant traction across various fields in recent years. Notably, it has demonstrated success in medical imaging analysis \cite{WAN20211,raghu2019transfusion, ISMAEL2021114054, MEHMOOD202143}. Its application in particle physics has been relatively limited, with examples including the classification of neutrino interactions \cite{Chappell2022}, the training of emulators for simulations of relativistic heavy ion collisions \cite{Liyanage2022}, enhancement of weak supervision searches \cite{Beauchesne2024}, and the development of efficient, data-efficient jet tags to capitalize on universality \cite{Dreyer2022} etc.\\ 

The structure of the paper is as follows: Section \ref{Transfer_Learning} introduces the Transfer Learning technique. Section \ref{Methodology} covers the methodology of this investigation, including data generation using AMPT and the preparation of inputs for the deep learning models. This section also addresses both the default and customized pre-trained models utilized in the study. Section \ref{result} discusses the training procedure, evaluation metrics, and model performance after fine-tuning. The effect of the bin size of input images on the the studied models' performance is also examined in this section. Finally, Section \ref{summary} outlines the key findings of the study.

\begin{figure*}[t!]
\centering 
\includegraphics[width=\linewidth]{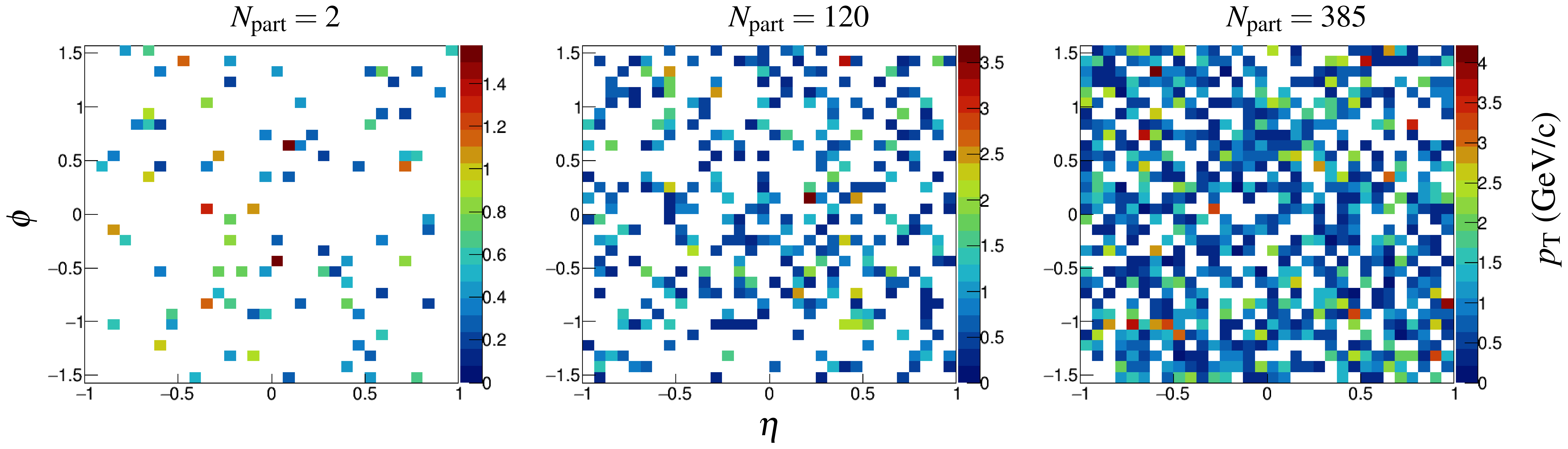}	
\caption{Two-dimensional (2D) $p_{\rm T}$-weighted $\left(\eta - \phi \right)$ spectra of charged hadrons generated using AMPT-SM for Au+Au collisions at $\sqrt{s}=200$ GeV for three different values of $N_{\mathrm{part}}$ with a binning of $32 \times 32$.} \label{fig_1}%
\end{figure*}

\section{Transfer Learning Method}
\label{Transfer_Learning}
Conventional machine learning and deep learning methods have been extensively used across several domains, yielding significant success in numerous instances \cite{rozenwald2020machine, Crawford2015, ALDULAIMI2019534,sym11020256}. Machine learning models, however, perform to the best of their ability when the feature space or data distribution of the source and target are the same. Consequently, the performance of the model degrades when the feature space or distribution of test data differs from that of train data. In other words, machine learning models are designed for specific tasks. As a result, any modifications to the feature space require the model to be rebuilt and retrained from scratch. Furthermore, training a new ML model, especially a deep learning model, is a time-intensive and resource-demanding task, requiring substantial amounts of data and processing power. When training data is limited, the performance of DL models deteriorates significantly. This is where transfer learning becomes handy. Transfer learning \cite{pan2010} refers to the method that reuses a previously trained model in a new task (called the target task) by leveraging the knowledge obtained from other related but different tasks (called source tasks). The core idea behind Transfer Learning involves utilizing the knowledge acquired by a model from a task with abundant training data and applying it to a new task with limited data. Rather than beginning from scratch, patterns acquired from previous tasks are used as a starting point for new tasks. The two major advantages of transfer learning are faster learning speed and the non-requirement of vast amounts of data. Generally, there are two ways to use a pre-trained model in transfer learning technique: fixed feature extraction and fine-tuning \cite{Yamashita2018}. In the fixed feature extraction method, the top layers, i.e., fully connected (FC) layers of the pre-trained models, are removed while the rest of the network, which comprises a series of convolution and pooling layers known as the convolutional base, serves as a fixed feature extractor. On top of the fixed feature extractor, any machine learning classifier can be put in addition to the standard FC layers. Training is limited to the additional layers on a particular dataset of interest. On the other hand, fine-tuning is an optional step, where after the feature extraction process, the entire or a part of the pre-trained network is retrained with the new data with a very low learning rate. This method can yield significant enhancement in the performance of the model by gradually adjusting the pre-trained characteristics to the new data. For this present work, both fixed feature extraction and fine-tuning methods have been implemented.

\section{Methodology}
\label{Methodology}
\subsection{Data Generation}
The input data for the present analysis is generated with the help of AMPT (a multi-phase transport) model. AMPT is a hybrid transport model extensively used to simulate relativistic heavy-ion collisions \cite{Lin:2005}. AMPT has four components, which include initialization of collisions by the heavy-ion jet interaction generator model (HIJING) \cite{Wang:1991}, parton transport by Zhang’s Parton Cascade model (ZPC) \cite{Zhang:1998}, hadronization process of the partons by quark coalescence model \cite{He:2017} in string melting version of AMPT and via the Lund string fragmentation scheme \cite{Andersson:1983} in the default version of AMPT. The last component of the AMPT model is the hadronic interactions based on the A Relativistic Transport (ART) model \cite{Li:1995}. In the current study, the string melting (SM) version of the AMPT model is used for generating Au+Au collisions at $\sqrt{s}=200$ GeV.

\begin{figure*}[t]
\centering 
\includegraphics[width=\textwidth, angle=0]{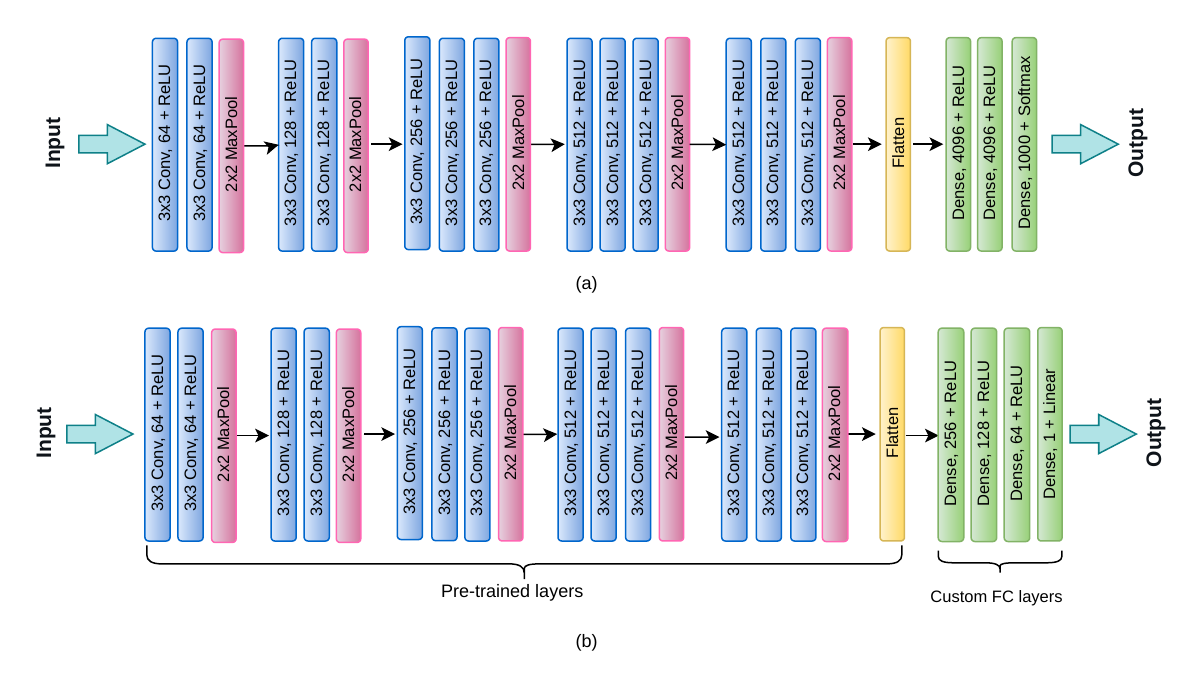}	
\caption{Architecture of the VGG16 model: (a) Original configuration \cite{vgg16}, (b) Customised for the proposed work.} \label{fig_VGG}%
\end{figure*}

\subsection{Input Preparation}
For conducting regression tasks using transfer learning effectively, it is crucial to carefully select input features that exhibit a robust correlation with the target variable. This involves identifying and incorporating features that contribute significantly to the predictive power of the model, thus ensuring its ability to generalize well to new data. The current investigation focuses on utilizing the STAR detector \cite{STAR:2002eio} kinematics from the RHIC experiment. The study incorporated three key raw observables of charged hadrons generated in Au+Au collisions at $\sqrt{s}=200$ GeV for input preparation: transverse momentum ($p_{\rm T}$), pseudorapidity($\eta$), and azimuthal angle ($\phi$). Transverse momentum weighted $\left(\eta - \phi \right)$ spectra were used as input to the DL models. Taking the realistic detector conditions into account, charged hadrons within the mid-rapidity ($\lvert \eta \rvert <1$) region and with a transverse momentum above 0.2 GeV/c were selected. These observables were transformed into two-dimensional (2D) $p_{\rm T}$-weighted $\left(\eta - \phi \right)$ histograms for each event with $32 \times 32$ bins. Each bin of the histograms is constructed by summing the $p_{\rm T}$ of particles within a specific range of pseudorapidity ($\eta$) and azimuthal angle ($\phi$). These ranges, denoted by $d\eta$ and $d\phi$, are fixed values (0.0625 and 0.098 units, respectively). Fig.~\ref{fig_1} illustrates the $p_T$-weighted $\left(\eta - \phi \right)$ spectra having $32 \times 32$ bins for different values of $N_{\rm part}$. Each images were resized to $32 \times 32$ pixels, and the image pixel values were normalized by scaling them to the range [0, 1]. Normalizing the data helps neural networks to learn more effectively and converge faster.

\begin{figure*}[t!]
\centering 
\includegraphics[width=\textwidth, angle=0]{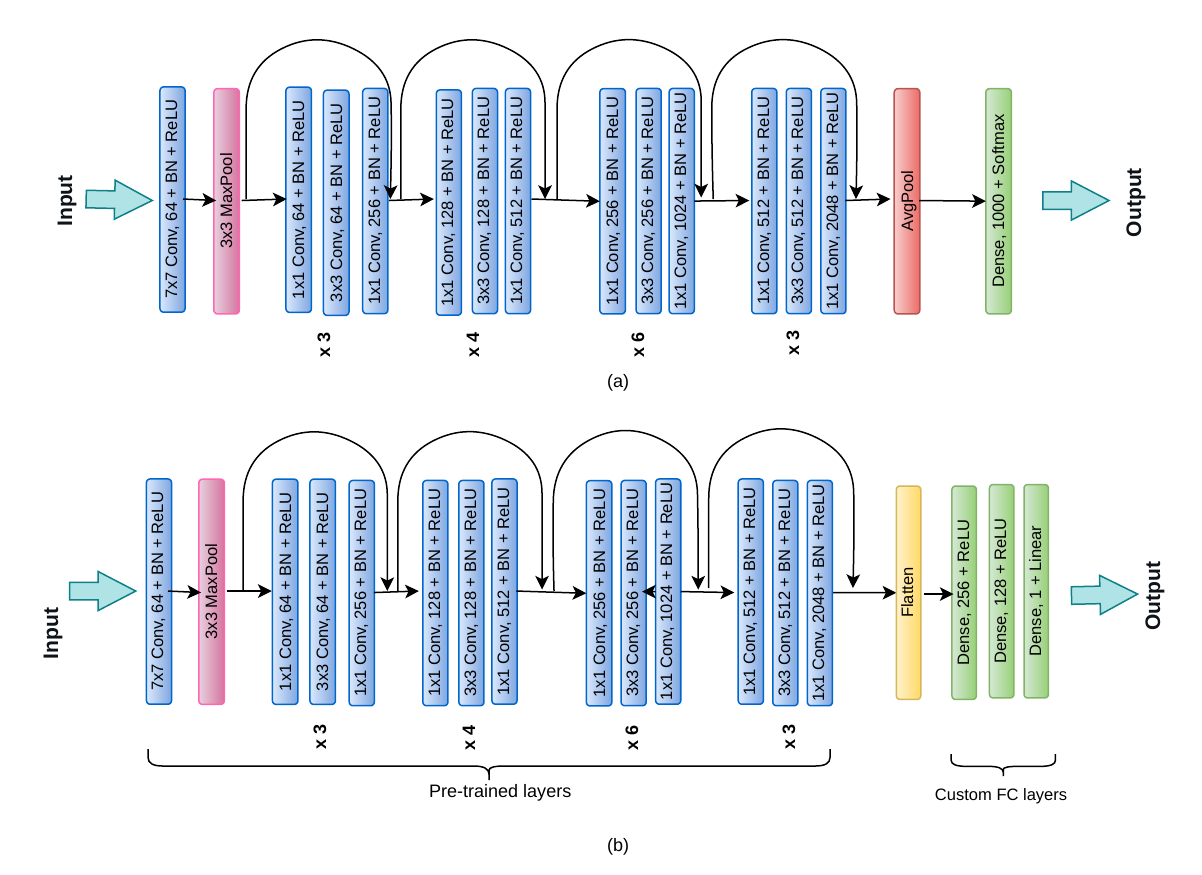}	
\caption{The architecture of the ResNet50 model: (a) Original configuration \cite{Resnet50}, (b) Modified version tailored for this study.}
\label{fig_ResNet}%
\end{figure*}

\subsection{Models Used}
In this study, we employed three established pre-trained architectures: VGG16 \cite{vgg16}, ResNet50 \cite{Resnet50}, DenseNet121 \cite{DenseNet}, known for their high performance in image classification tasks on the ImageNet dataset \cite{ImageNet}. ImageNet is a widely utilized dataset in Computer Vision, comprising 1.28 million natural images categorized into 1,000 classes. It has been instrumental in the progress of both developing and evaluating computer vision models.

\subsubsection{VGG16}
The VGG16 model \cite{vgg16} is a convolutional neural network architecture that gained prominence for its simplicity and effectiveness in image recognition tasks. Developed by the Visual Geometry Group (VGG) at the University of Oxford, VGG16 is characterized by its deep architecture consisting of 16 layers, including 13 convolutional layers and 3 fully connected layers. The VGG model won the ImageNet 2014 competition \cite{ILSVRC15} with a staggering accuracy of 92.7\% on the ImageNet dataset. The VGG16 variant of the architecture consists of five blocks of convolutional layers followed by three fully connected layers. There are a total of 13 convolution layers, each with a kernel size of $3 \times 3$  with a stride of 1 and padding of 1 to ensure that every activation map keeps the same spatial dimensions as its preceding layer. Max pooling layers with $2\times 2$ kernels with a stride of 2 and no padding are used at the end of each convolutional block to down-sample the spatial dimension. Rectified Linear Unit (ReLU) activation function \cite{Nair:2010} is applied in all layers except the last one where softmax activation function \cite{softmax} is used. It is to be mentioned that while VGG16 was primarily designed for image classification tasks, it can also be repurposed for regression tasks. However, its architecture, consisting of convolutional and fully connected layers, was originally optimized for classification tasks and may not be ideal for regression tasks. In this investigation, we replaced the final three fully connected layers with our custom layers, consisting of four fully connected layers with 256, 128, 64, and 1 neurons, respectively, as shown in Fig.~\ref{fig_VGG}(b).

\subsubsection{ResNet50}
To enhance the performance of deep learning networks, a common strategy is to increase their depth. However, deeper networks often suffer from the vanishing gradient problem, leading to degraded performance. K. He et al. \cite{Resnet50} introduced the Residual Network (ResNet) to address this issue. ResNet solves the problem of the vanishing gradient problem even with extremely deep neural networks. In residual learning networks, blocks of convolutional layers are skipped by introducing shortcut connections between the input and output of a stack of a few convolution layers to form blocks named residual blocks. Shortcut connections allow the models to be made deeper without degrading performance. ResNet comes in several architectures, with the 50-layer version utilized in this study. This architecture comprises 50 layers (49 Convolution layers and one fully connected layer) with 23,534,592 trainable parameters. For this study, the ResNet50 structure was modified by replacing the fully connected layer with three fully connected layers with 256, 128, and 1 neurons, respectively. Both the original and modified ResNet50 structure are depicted in figure \ref{fig_ResNet}. 

\begin{figure*}[h]
\centering 
\includegraphics[width=\textwidth, angle=0]{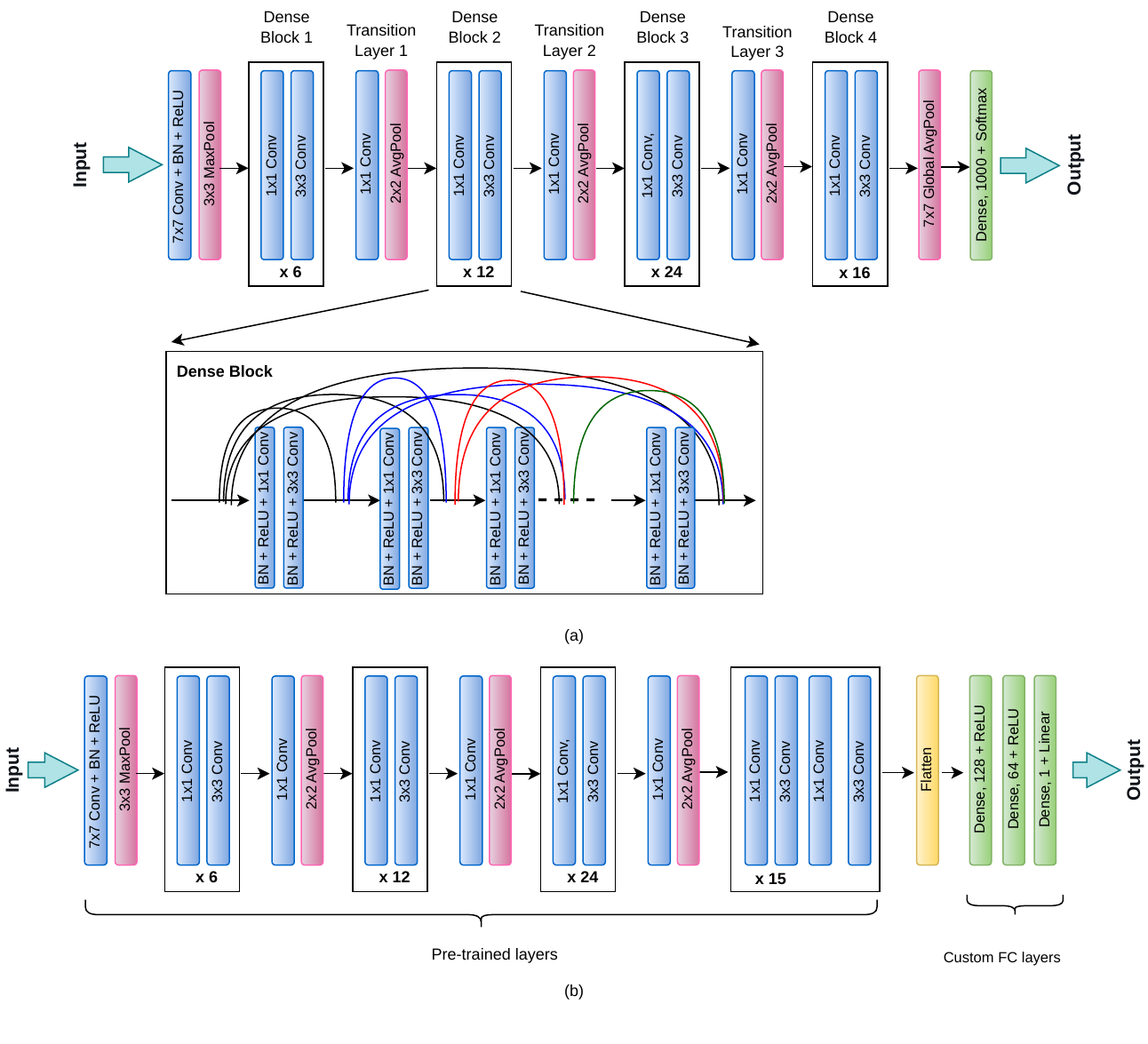}	
\caption{Architecture of DenseNet121 model: (a) Original configuration (b) Customised version for this study.} 
\label{fig_DenseNet}%
\end{figure*}

\subsubsection{DenseNet121}
The Dense Convolutional Network (DenseNet) proposed by Huang et al. in 2017 \cite{DenseNet} is a convolutional neural network with a densely connected structure. In DenseNet, each layer receives inputs from all preceding layers and passes its own feature maps to all subsequent layers. Consequently, the final output layer has direct access to information from every prior layer. This distinctive structure enhances the learning ability of the network by reusing features, which notably reduces the number of parameters and mitigates the vanishing gradient problem. A DenseNet architecture begins with a basic convolution with a kernel size of $7\times7$  followed by a max pooling layer with $3\times3$ kernels with a stride of 2. The DenseNet architecture consists of several dense blocks and transition layers in series. Each dense block is composed of a varying number of convolutional blocks which follow a specific sequence. Each convolutional block starts with a batch normalization layer, then a ReLU activation function, and finally, a Conv2D layer. Transition layers are placed in between two dense blocks. Each transition layer includes a $1\times 1$ convolutional layer, followed by a $2 \times 2$ average pooling layer with a stride of 2. Finally, a global average pooling layer is followed by a classification layer (fully connected layer with softmax activation function). We have modified the DenseNet121 architecture for our study by replacing the global average pooling layer with a flatten layer to prepare the data for the fully connected layers. The final classification layer is then substituted with three new fully connected layers, featuring 256, 128, and 1 neurons, respectively, as shown in Fig.~\ref{fig_DenseNet}. This modification was tailored for regression rather than classification.

\begin{figure*}[t!]
\centering 
\includegraphics[width=1\linewidth, angle=0]{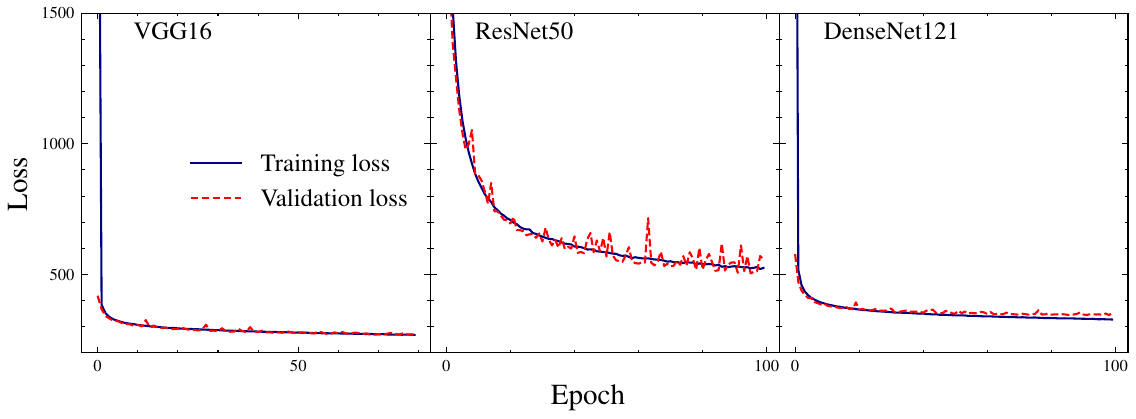}	
\caption{Mean-Squared-Error Loss per epoch during the training of all the transfer learning models with 150K minimum biased Au+Au collisions at $\sqrt{s}=200$ GeV.} 
\label{Learning_Curve}%
\end{figure*}

\section{Results and Discussion}
\label{result}
Although initially trained on classification tasks, the transfer learning models used in the current study are adapted for regression, a new task requiring them to predict continuous values rather than categories. Moreover, the image dataset used for pre-training the original models differs from our heavy-ion dataset. We started with the fixed feature extraction approach of Transfer learning. The top layers of the models were replaced with custom fully connected layers forming the regression head. Each model now consists of two parts: the convolutional base and the regression head. The convolutional base, which is the pre-trained model, extracts domain-independent features from the input images. These extracted features are then transferred to the custom fully connected layers, which utilize them to perform regression on the input images based on the training data. During training, we started with freezing all the layers of the convolutional base and only trained the regression head. Freezing prevents weights in those layers from being modified during training. The weights of all the pre-trained models were initialized with the ImageNet dataset.

For all networks, Adam optimizer \cite{Adam:2015} was used with a learning rate of 0.0001. To prevent over-fitting, (i) L1 regularization \cite{Tibshirani:1996} and L2 regularization \cite{Hinton:1986} and (iii) EarlyStopping technique \cite{EarlyStop:2007} were employed. 
Each model was trained for up to 100 epochs with a batch size of 64, and the Mean-Squared-Error (MSE) loss function was used to monitor model performance during training. The EarlyStopping technique helps to prevent overfitting by monitoring the validation loss with a patience level of 10; if overfitting occurs, the weights with the minimum loss were selected. ReLU was used as the activation function in all fully connected layers of the regression head, except for the last layer, where a linear activation function was used. All the model training and evaluation tasks in this study were performed on a system with Intel(R) Core(TM) i5-10300H CPU @ 2.50GHz $\times 8$ and 16 GB DDR4 RAM.

\begin{figure*}[h]
\centering 
\includegraphics[width=1\linewidth, angle=0]{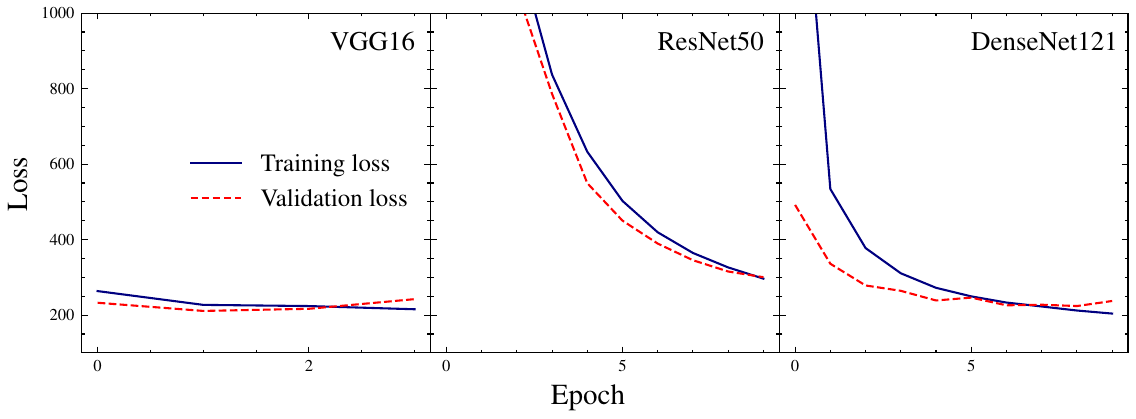}	
\caption{Mean-Squared-Error Loss per epoch during the fine-tuning of all the models with learning rate of 0.00001 }.
\label{Learning_Curve_FT}%
\end{figure*}

\begin{figure*}[h]
\centering 
\includegraphics[width=1\linewidth, angle=0]{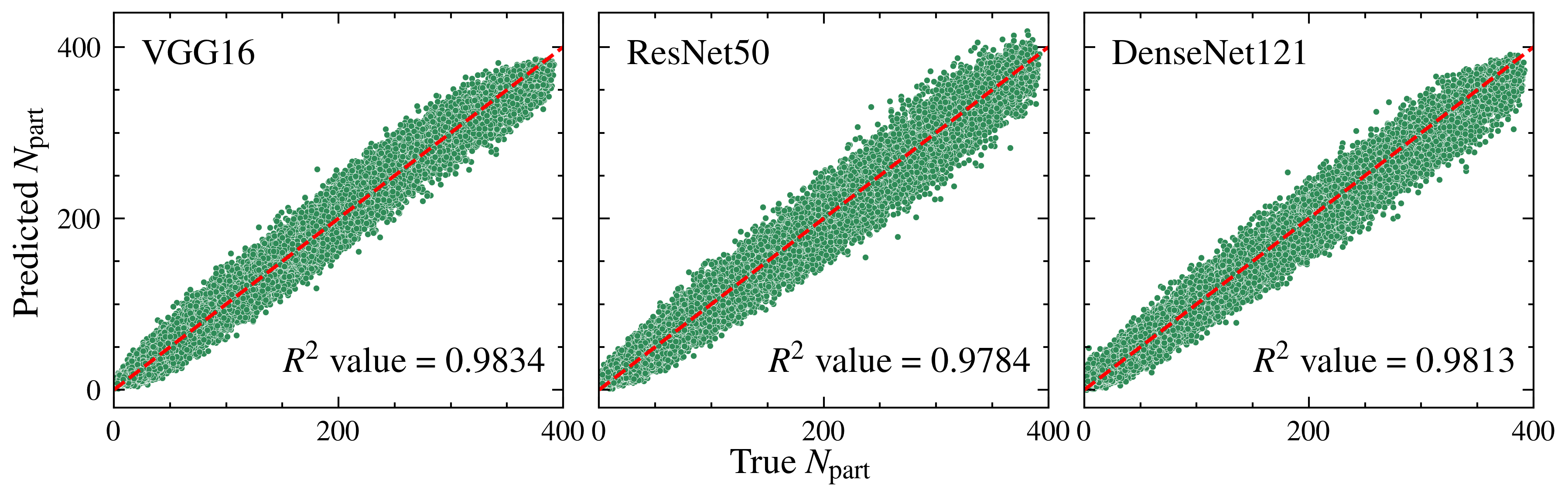}	
\caption{Predicted values by the fine-tuned models versus actual values of $N_{\rm part}$ for 30K AMPT-SM simulated minimum biased Au+Au collisions at $\sqrt{s}=200$ GeV. The red dashed line indicates perfect agreement between predicted and actual values.} 
\label{fig_corelation}%
\end{figure*}

\subsection{Evaluation Metrics}
To check the performance of the models, we used two evaluation metrics, namely the R-squared ($R^2$) value and the Mean-Squared-Logarithmic-Error (MSLE). $R^2$ value or the coefficient of determination is a statistical measure in a regression model that shows how well the model predicts the outcome of observed data. $R^2$ values range from 0 to 1 where 1 represents a perfect fit. The $R^2$ value is expressed as
\begin{equation}
R^2 = 1 - \frac{\sum_{i=1}^{N}{\left(y_i^t-y_i^p\right)}^2}{\sum_{i=1}^{N}{\left(y_i^t- \langle y^t \rangle \right)}^2}.
\label{eq1}
\end{equation}
where $y_i^t$, $y_i^p$ and $\langle y^t \rangle$ are the true values, model predicted values and the mean of true values of target variable respectively.
Considering the large range of target variable ($N_{\rm part}$ in our case), we used MSLE value as the additional performance evaluator as it is useful when the data has a wide range of values. MSLE gives the average of the squared differences between the logarithms of the predicted and true values  and is defined as
\begin{equation}
\text{MSLE} = \frac{1}{N}\sum_{i=1}^{N} {\{ \log \left(1+y_i^{t}\right)-\log\left (1+y_i^{p}\right)\}}^2.
\label{eq2}
\end{equation}

\subsection{Training of the models}
Using the AMPT-SM setup, a total of 180K Au+Au collisions at $\sqrt{s}=200$ GeV were generated. We used 150K events to train the models, with 75\% of these events allocated for training and 25\% for validation. Figure \ref{Learning_Curve} shows the MSE loss curve for both training and validation as a function of the epoch number. The figure clearly demonstrates that the training and validation loss for each model considered in this study decreases significantly with the number of epochs. After a few epochs, both the training and validation losses align closely, indicating that the trained models generalize well when using the validation dataset to predict the target variable. The performance of the models is summarized in Table~\ref{Performance_FE}, which lists the MSLE and $R^2$ values. The results show that all models perform satisfactorily, with VGG16 outperforming the others.

\begin{table}[h]
\centering
\caption{Performance of the models after feature extraction}\label{Performance_FE}%
\begin{tabularx}{\linewidth}{X X c}
 \hline
 Model & MSLE & $R^2$  \\  
\hline
VGG16 & 	0.0800 & 0.9785 \\ 
ResNet50 & 0.1645 & 0.9552 \\ 
DenseNet121 & 0.1037 & 0.9714 \\ 
\hline
\end{tabularx}
\end{table}

\begin{figure*}[h]
\centering 
\includegraphics[width=1\linewidth, angle=0]{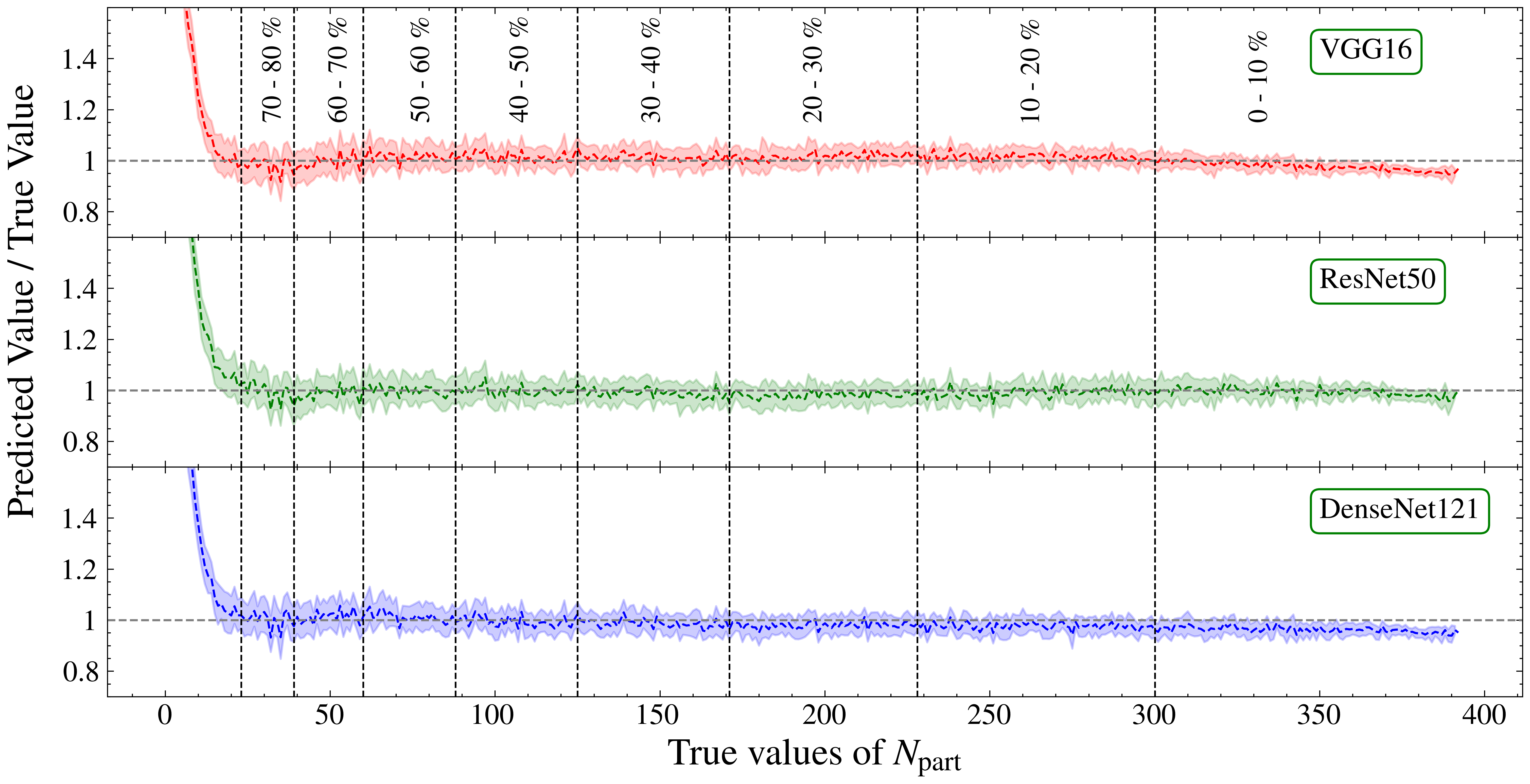}	
\caption{The ratio of predicted and true $N_{\rm part}$ as a function of true $N_{\rm part}$ for all the fine-tuned models.} 
\label{fig_ratio}%
\end{figure*}

\subsection{Fine Tuning}
For better performance, all the trained models were retrained using the fine-tuning technique. In the fine-tuning approach, all the layers of the convolutional base of the pre-trained model are unfrozen, and the entire model is trained again with a reduced learning rate. The low learning rate is essential as otherwise, it may impact the previously learned knowledge. All the models were thus retrained with a learning rate of 0.00001 for 10 epochs. As the unfrozen models are much larger, there is always a possibility of over-fitting. The EarlyStopping technique was applied with a patient level of 2 to address overfitting. When overfitting was detected, the model weights were restored with minimum validation loss. Figure \ref{Learning_Curve_FT} displays the learning curves of the models during fine-tuning, indicating that both the training and validation losses continue to decrease. This further reduction in loss demonstrates the effectiveness of the fine-tuning process. It also suggests that the models are increasingly able to generalize from the training data to the validation data, improving their predictive performance. These results highlight the robustness of the models in adapting to the new regression task. Table \ref{Performance_FT} shows the performance of the fine-tuned models compared with our previous CNN model from \cite{Basak2023}. It indicates that the performance of all models improved after fine-tuning, with VGG16 emerging as the top performer, matching the performance of our previous CNN model \cite{Basak2023}.

\begin{table}[h]
\centering
\caption{Performance of the models after fine tuning}
\label{Performance_FT}%
\begin{tabularx}{\linewidth}{X X c}
\hline
 Model & MSLE & $R^2$  \\   
\hline
VGG16 & 	0.0602 & 0.9834 \\ 
ResNet50 & 0.0754 & 0.9784 \\ 
DenseNet121 & 0.0767 & 0.9820 \\ 
CNN \cite{Basak2023} & 0.0592 & 0.9847\\ 
\hline
\end{tabularx}
\end{table}

Figure \ref{fig_corelation} presents the correlation plot between the predicted value of $N_{\rm part}$ and the ground truth values of $N_{\rm part}$ for all three transfer learning models evaluated in this study. The plot clearly illustrates that the predicted values are densely clustered around the diagonal dashed line, signifying that the predicted values closely match the real values. This high level of accuracy is consistent across all three transfer learning models, demonstrating the robustness and reliability of this approach for this task.

Furthermore, we have plotted the ratio of predicted to true $N_{\rm part}$ values against the true $N_{\rm part}$ for all three fine-tuned models in Figure~\ref{fig_ratio}. The figure shows that all models perform better for semi-central and central collisions. The poorer performance of all models for peripheral collisions can be attributed to the lower number of particles produced in these collisions, resulting in less information being fed to the deep-learning models for training. However,  the VGG16 and DenseNet121 models also show a slight decline in performance for extreme central collisions. It is to be noted that the extreme central collisions are the rare event present in the dataset. This rarity leads to data imbalance, adversely affecting the performance of these underrepresented extreme central collisions.

\begin{figure*}[h]
\centering 
\includegraphics[width=\linewidth]{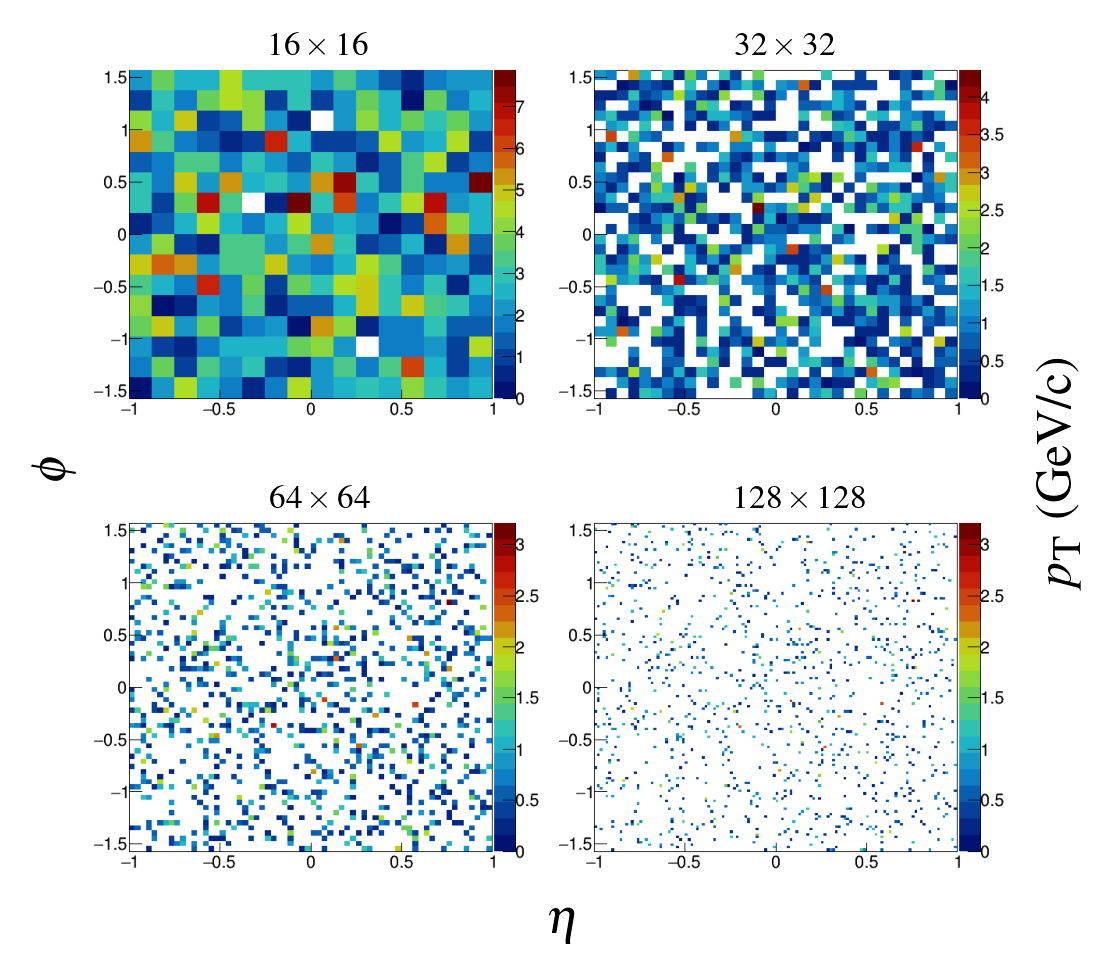}	
\caption{Two-dimensional (2D) $p_T$ weighted $\left(\eta - \phi \right)$ spectra drawn for charged hadrons generated using AMPT-SM for a single Au+Au collision at $\sqrt{s}=200$ GeV with $N_{\mathrm{part}}=390$ for different bin sizes.} \label{fig_input_bin_size}%
\end{figure*}

\subsection{Effect of the bin-size of input histograms}
In the present regression task, $p_T$-weighted ($\eta-\phi$) histograms with $32\times 32$ bins were employed as input for the transfer learning models. The size of histogram bins can significantly influence the outcome of a deep learning (DL) task. For instance, smaller bin sizes capture finer details in the distribution of pixel intensities, which, in turn, can help the model to learn subtle variations in the data, potentially leading to better performance. However, smaller bins can also introduce noise if these fine details are irrelevant, leading to increased statistical fluctuations and reduced regression accuracy. Additionally, finer bin sizes may lead to overfitting, where the model learns noise rather than the true pattern, reducing its generalizability to new data. Conversely, a broader bin size reduces noise and highlights broader patterns in the data. However, if the binning is too coarse, it might miss important variations in the data, leading to poorer performance. Therefore, striking a balance between resolution and statistical reliability is paramount. This study investigates the effect of varying histogram bin sizes on the performance of centrality determination in heavy-ion collision. This study used the best-performing fine-tuned VGG16 model for different histogram bin configurations such as $16 \times 16$, $32 \times 32$, $64 \times 64$, and $128 \times 128$. Figure~\ref{fig_input_bin_size} shows the $p_T$-weighted ($\eta-\phi$) histograms of charged hadrons produced in Au+Au collision with $N_{\rm part}=390$ for different bin configurations. It is to be noted that for this present study, uniform image size ($32 \times 32$) for all four configurations was chosen during the training procedure. Fixed image size ensures that the number of parameters of the DL model remains the same. Table \ref{Performance_bin_size} shows the performance of the fine-tuned VGG model with different histogram bin configurations, evaluated using metrics such as MSLE and $R^2$. The results indicate that histograms with $64\times 64$ bin configurations performed better than other bin sizes.  

\begin{table}[h]
\centering 
\caption{Performance of fine-tuned VGG16 model for different bin sizes of input histograms.}\label{Performance_bin_size}%
\begin{tabularx}{\linewidth}{X X c}
\hline\noalign{\smallskip}
No. of bins & MSLE & $R^2$  \\   
\noalign{\smallskip}\hline\noalign{\smallskip}
$16 \times 16$ & 	0.0678 & 0.9760 \\ 
$32 \times 32$ & 0.0602 & 0.9834 \\ 
$64 \times 64$ & 0.0576 & 0.9852 \\ 
$128 \times 128$ & 0.0806 & 0.9816 \\ 
\noalign{\smallskip}\hline
\end{tabularx}
\end{table}

\begin{figure*}[h]
\centering 
\includegraphics[width=1\linewidth, angle=0]{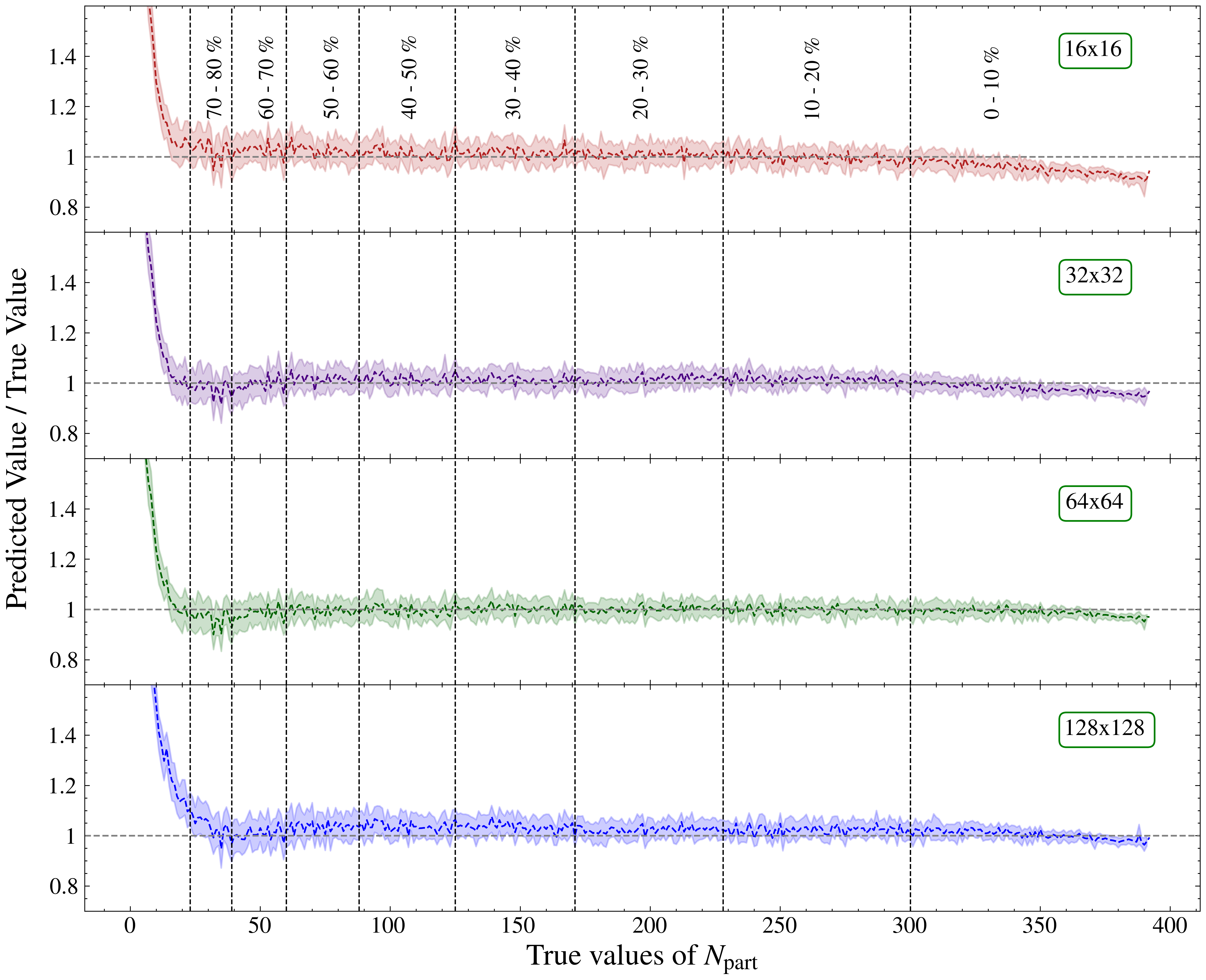}	
\caption{Ratio of predicted and true values of $N_{\rm part}$ as a function of true $N_{\rm part}$ for different input histogram bin configurations. The vertical slices represent the centrality percentiles.} 
\label{fig_ratio_bin}%
\end{figure*}

In Fig. \ref{fig_ratio_bin}, the ratio of the predicted and true $N_{\rm part}$ values are plotted as a function of true $N_{\rm part}$ values for different bin configurations. This figure reveals a similar decline in performance for peripheral events across all studied configurations. Additionally, a slight decrease in performance is observed for the most central collisions in nearly all studied bin configurations. To quantitatively assess model performance for central collisions, MLSE values for all configurations were calculated for the 0-10\% centrality class and presented in Table~\ref{Performance_bin_size_centrality}. This study indicates that for the most central collisions, model performance improves as the number of bins or the image resolution increases. Smaller bin sizes in the input histograms enhance granularity and provide a more detailed representation of the input data, thereby improving the model performance.

\begin{table}[h]
\caption{Performance of the pre-trained VGG16 model in terms of MSLE values for $0- 10 \%$ central Au+Au collisions at $\sqrt{s}=200$ GeV when trained and tested with images having different bin sizes.}\label{Performance_bin_size_centrality}%
\begin{tabularx}{\linewidth}{X X X X X}
\hline\noalign{\smallskip}
 No. of bins & $16 \times 16$ & $32 \times 32$ & $64 \times 64$ &   $128 \times 128$ \\   
 \noalign{\smallskip}\hline\noalign{\smallskip}
MSLE & 0.0066 & 0.0030 & 0.0027  & 0.0025 \\ 
\noalign{\smallskip}\hline
\end{tabularx}
\end{table}

\section{Summary and outlook}
\label{summary}
In this investigation, the transfer learning technique is used to determine collision centrality in terms of the number of participants ($N_{\rm part}$) for Au+Au collisions at $\sqrt{s}=200$ GeV. For this purpose, three popular pre-trained models, namely VGG16, ResNet50, and DenseNet121, were deployed. We customized the classification heads of all three models to meet the requirements of this regression task while keeping the convolutional heads intact. Simulated data generated using the string melting version of the AMPT model was used to train the pre-trained models. A total of 150K events were used for the training and validation of the models, where 75\% of the total events were used for training and the remaining 25\% were used for validation. The models were first used for feature extraction from the input histograms, where only the classification heads were trained, keeping the pre-trained weights of the convolutional heads frozen. Next, we fine-tuned the entire models with a lower learning rate of 0.00001 to adapt the pre-trained weights to the new data. All three models achieved good accuracy in predicting $N_{\rm part}$ values, especially for semi-central and central collisions, with the VGG16 model outperforming the others. The performance of the VGG16 model is comparable to that of the CNN model \cite{Basak2023}, with MSLE and $R^2$ values of 0.06 and 0.98, respectively. This study suggests that the transfer learning technique can effectively determine centrality, even when the model is pre-trained on images from different domains. Additionally, we investigated the impact of input histogram bin size on model performance by training and testing the VGG16 model with histograms of varying bin sizes while keeping the image pixel size constant during training. It was observed the bin size of the histogram has an effect on the performance of the model, where histograms with $64 \times 64$ bin sizes performed better than other bin sizes. It was also found that the accuracy of determination of $N_{\rm part}$ increases with the decrease of the bin size for central collisions.

This study highlights the usefulness and potential of transfer learning techniques in heavy-ion physics data analysis. By leveraging pre-trained models, these techniques can be applied to determine various important observables, opening new avenues for research in this field. Transfer learning thus holds promise for enhancing our understanding of heavy-ion collisions.


\bibliographystyle{spphys}       



\end{document}